\documentstyle[12pt,aasms4]{article}
\begin{document}
 
\title{Deconstructing A3266: A Major Merger in a Quiet Cluster}

\author{R. A. Flores}
\affil{Department of Physics and Astronomy,
University of Missouri--St.Louis,\\
St.Louis, MO 63121-4499\\
Electronic mail: Ricardo.Flores@umsl.edu}

\author{H. Quintana\altaffilmark{1,2}}
\affil{Department of Astronomy and Astrophysics,\\
P. Universidad Catolica de Chile, Casilla 306, Santiago 22 , Chile\\
Electronic mail: hquintana@astro.puc.cl}

\author{M. J. Way}
\affil{Department of Astrophysical Sciences,\\
Princeton University, Princeton, NJ 08544\\
Electronic mail: mway@astro.princeton.edu}

\altaffiltext{1}{Presidential Chair in Science 1995, 1999}

%%%%%%%%%%%%%%%%%%%%%%%%%%%%%%%%%%%%%%%%%%%%%%%%%%%%%%%%%%%%%%%%%%%%%%

\begin{abstract}

We present results of simple N-body simulations that strengthen the suggestion
that A3266 is
composed of two subunits of comparable mass that have recently merged. Both
the real cluster and the N-body dark-matter cluster show mixed signals
of substructure under statistical tests. However, in a decidedly
non-statistical approach allowed by the wide-area coverage and large number of
redshifts they measured in A3266, Quintana, Ram\'{\i}rez, \& Way (1996; QRW)
{\it sliced the real cluster in redshift space} to uncover a peculiar spatial
distribution of galaxies that they suggested was the result of a recent
merger. In our simulations, a similar distribution is the result of an ongoing
merger between two comparable-mass units that started about $2\times10^9$
years ago in the N-body simulations. We also find that the distribution of
emission line galaxies in A3266 traces the same structure. We discuss further
tests of our merger
hypothesis, and speculate on the possibility that a similar process might be
occurring in other, apparently-relaxed clusters at the present epoch.

\end{abstract}

%%%%%%%%%%%%%%%%%%%%%%%%%%%%%%%%%%%%%%%%%%%%%%%%%%%%%%%%%%%%%%%%%%%%%%


%%%%%%%%%%%%%%%%%%%%%%%%%%%%%%%%%%%%%%%%%%%%%%%%%%%%%%%%%%%%%%%%%%%%%%

\section{Introduction}
	There have been numerous analyses of substructure in clusters of
galaxies over the past decade (for a review, see e.g. \cite{W94}), partly
motivated by the expectation that its study will help unravel the cosmogony
underlying their formation. For example, the presence of substructure can teach
us about the overall matter density in the universe (\cite{RLT92}), although
subject to the uncertainty in the rate at which substructure is erased
(\cite{KW93}; \cite{LC93}). Another example is the spatial distribution of
substructure, which can give clues to the formation process (\cite{WJF95}). 

	Most studies of substructure have been statistical in nature, with
both optical and X-ray studies suggesting that $30\% - 50\%$ of clusters show
evidence of substructure in their galaxy and/or gas distribution (\cite{GB82};
\cite{DS88}; \cite{JF92}; \cite{SSG93}; \cite{GEFGMM97}; \cite{SSG98}). A
sharper
agreement on what fraction of clusters shows substructure in its projected
distribution is precluded partly because many features are generically
considered as substructure (see a discussion in \cite{GSSA98}). In addition
the fraction appears to be much larger ($\sim 80\%$, \cite{B94}) when evidence
for substructure is also looked for in the form of multimodal velocity
distributions. This kind of test is more sensitive to substructure arising from
line-of-sight mergers in clusters (\cite{PRBB96}), which can then
appear fairly smooth in their projected distribution. In fact, one is tempted
to ask if the remaining $\sim 20\%$ might not be clusters where a line-of-sight
merger has comparable-mass subclumps that have substantially decelerated after
the cores collided. These would be difficult to uncover in the
projected distribution of galaxies in the cluster due to orientation, and in
the velocity distribution because the subclumps have nearly stopped.
Our analysis of A3266 here suggests that it is an example of such clusters.

	Studies have also been directed at individual clusters, such
as the Coma cluster (see e.g. \cite{FW87}; \cite{DM93}; \cite{WBH93};
\cite{BRLK94}; \cite{CD96}; and references therein). Other examples include
A400 (\cite{BGHFJB92}), A2634 (\cite{PRBHOBH93}), and a recent study of the
A3558 cluster complex (\cite{BPRZZ98}). In these studies one attempts to go
beyond simply establishing that there is evidence of substructure and into
modeling the possible dynamics that gives rise to the observed structure.
The study of substructure in individual clusters can be helpful to check
trends expected in cosmological models. For example, it appears
that an absence of cooling flows occurs in clusters undergoing a merger (see
\cite{BRPPOV95}, and references therein), as expected from hydro/N-body
simulations (\cite{RBL93}). Also, these studies can help unravel whether the
substructure is the result of a major merger or an aggregate of accreeted
small units. This question has also been addressed statistically by
Gonz\'{a}lez-Casado, Salvador-Sol\'{e}, Serna \& Alimi (1998).

	Here we present an analysis of spectroscopic data for galaxies in the
ACO galaxy cluster A3266, obtained by QRW, which
we interpret by means of simple N-body simulations to infer the dynamical
state of the cluster. We briefly summarize the observations
of QRW in the next section, together with further analysis of the data.
We then present simple, 1000-particle N-body simulations of the cluster. We
find that simple statistical tests give mixed signals about the presence of
substructure in the simulated cluster, much like we find for the real cluster
in section 2, and despite the fact that there is an ongoing merger in the
N-body cluster.  Finally, we close with a section of discussion of this
analysis and the conclusions we draw from it.

%%%%%%%%%%%%%%%%%%%%%%%%%%%%%%%%%%%%%%%%%%%%%%%%%%%%%%%%%%%%%%%%%%%%%%

\section{The Observations, and Further Analysis of the Data}

	QRW compiled a total of 387 velocities in an area
approximately $1.8\arcdeg\times1.8\arcdeg$ centered on A3266, most of
which (229)
were new velocities obtained from runs at the Cerro Tololo (CTIO) and Las
Campanas (LCO) observatories. A total of 317 galaxies were identified as
cluster members from the distribution of velocities, making it one of the
largest data sets of its kind for clusters in the ACO catalog.

	The first run was carried out using the Argus spectrograph at the
CTIO 4 m telescope during the early parts of two nights on 1990 February
17-19. Three Argus fields were observed in A3266 during this run, securing
new spectra for 46 objects. Three 900 s exposures for the
first field were taken on the second night of the run. Two 1800 s exposures
for the second field and two 1500 s for the third field were taken on the
final night. Given the measured stability of the instrument, a single long
exposure of the He-Ar comparison lamp was taken every
night to calibrate all exposures in wavelength.
To compare and check on possible zero-point shifts, several velocities of
standard stars and some galaxies with well-known velocities were measured
using one of the fibers. Exposures of a white spot in the dome and sky flats
were used to correct for pixel-to-pixel, large scale, and illumination
variations in the detector (the $800\times800$ pixels TI\#2 CCD) response.
Finally, the grating used was KLGL2, tilted to provide a wavelength coverage
from $\sim 3900-5600$ {\AA}. The preflashed CCD exposures were binned
$2\times1$
in the fibers-slit direction, giving a dispersion of 2.2 {\AA}/pixel
with a FWHM resolution of $\sim 8$ {\AA}.

	The whole $1.8\arcdeg\times1.8\arcdeg$ field around A3266 was
explored in the second run, using Shectman's fiber spectrograph mounted on
the 2.5m DuPont telescope at LCO. The run was carried out on the nights of
1990 October 22-25, and a total of 263 new spectra were obtained. Five
fields were used to cover (with considerable overlap) an area of approximately
$1.8\arcdeg\times1.8\arcdeg$, with exposure times adjusted between 80 and
120 min depending on the brightness of the selected galaxies in each of
the exposures. Standard quartz lamp exposures of the dome were used to
approximately correct for pixel-to-pixel variations of the 2D-Frutti
detector coupled to the spectrograph, but no corrections were made for the
(small) dark current in the detector. He-Ne comparison lamp exposures
were taken for wavelength calibration before and after each exposure.
With a 600 line/mm grating plus the 2D-Frutti detector, spectra covered the
range $\sim 3500-6900$ {\AA} with a dispersion of $\sim 2.6$ {\AA}/pixel,
and a resolution of $\sim 10$ {\AA}.

	A careful analysis was carried out by QRW in order to
combine the measurements from the two runs with previous data (mostly from
\cite{QR90} and \cite{TCG90}) and create the homogeneous, large catalog of
velocities we use here. We first consider the velocity distribution for
the 317 members identified in A3266. We find that even with this large number
of velocities there is no conclusive evidence of non-normality in the
velocity distribution, which would be indicative of the presence of
substructure. For example, the skewness
and kurtosis are .105 and 3.32 respectively. For a gaussian distribution,
values as high as these would occur 21.7\% and 9.64\% of cases respectively
(the mean and dispersion are estimated from the data). The Kolmogorov-Smirnov
(KS) statistic, however, appears to exclude the gaussian hypothesis at a much
higher confidence level (CL) of 95\% (see QRW). This is rather
surprising, since one would expect order statistic tests to be less sensitive
(\cite{BB93}). The KS test was not considered in \cite{BB93}, but it is easy
to carry out the analysis for it. In Table 1 we present the fraction of
times the KS, skewness and kurtosis tests would reject a Tukey distribution
(see \cite{BB93}, and references therein) as non-gaussian at the 95\% CL. It
can be seen there that the KS test never outperforms a combined skewness-and-
kurtosis test. In fact, skewness and kurtosis alone were used in the recent
analysis of the ENACS clusters (\cite{SSG98}), where A3266 would not have been
considered as having a non-gaussian velocity distribution.

{\scriptsize
\begin{deluxetable}{lcccccc}
\tablenum{1}
\tablecaption{Power of Tests}
\tablehead{
\colhead{Test}  &  \multicolumn{6}{c}{Tukey distribution parameters (g,h)} \\
\cline{2-7} \\
\colhead{}  & \colhead{(0.1,0)}   & \colhead{(0.2,0)} &
\colhead{(0,0.1)} & \colhead{(0,0.2)} & \colhead{(0.1,0.1)} &
\colhead{(0.2,0.2)}}
\startdata
KS       & 0.23 & 0.73 & 0.53 & 0.98 & 0.68 & 0.99\nl
skewness & 0.67 & 0.99 & 0.24 & 0.37 & 0.70 & 0.90\nl
kurtosis & 0.18 & 0.52 & 0.95 & 1.00 & 0.96 & 1.00\nl
\enddata
\end{deluxetable}
}

	The 5\% of samples with the highest values of
the KS statistic, and drawn from a gaussian distribution, are biased to high
skewness and kurtosis. Values such as those above would occur a fraction of
40.5\% and 18.7\% of the time (respectively) in that subset of samples.
Therefore, the QRW sample might just be an ``unlucky'' sample out
of a gausian distribution. After all, 5\% is not such a
low probability if we bear in mind that confidence levels refer to
any set of measurements, not just measurements in A3266. Moreover,
we find that if we assume {\it true} mean and dispersion of
$\sim 17830$ km/sec and $\sim 1190$ km/sec respectively, which are values
well within the measurement errors, all three tests give rather large
probabilities ($\sim 30\%, \sim 35\%$ and $\sim 46\%$ respectively for the
KS, skewness and kurtosis tests) that the set is drawn from a gaussian
parent. Thus, we conclude that the gaussian hypothesis cannot be excluded
with sufficient confidence.

	The distribution of the member galaxies in the plane of the sky can
be combined with the velocity information in order to further search for
departures from equilibrium. We find that the central region of A3266 is
entirely consistent with a spherical isothermal distribution. The cumulative
distribution of right ascensions (RA) or declinations (DEC) needed to perform
a KS test can be readily worked out for an isothermal distribution. We find
that the distribution of RA, $F(x)$, inside a box of size $2a$ ($2b$) in RA
(DEC) centered on the cluster must be
\begin{equation}
F(x) = {asinh^{-1}(b/a)+bsinh^{-1}(a/b)+xsinh^{-1}(b/x)+bsinh^{-1}(x/b)
\over 2(asinh^{-1}(b/a)+bsinh^{-1}(a/b))}\ ; x > 0
\end{equation}
and $F(x) = 1-F(|x|)$ for $x < 0$. (We have assumed that the galaxies trace the
mass and the cluster core radius is very small, as indicated by gravitational
lensing studies of rich clusters. See e.g. \cite{TKD99}). Since the galaxies
trace the mass, we take the cluster center to be the center of a smoothed
galaxy density map, shown in Fig. 1(a). A typical Monte Carlo (MC) realization
of the data in a window of the same size as in Fig. 1(a), using eq.(1), is
shown in Fig. 1(b). In the
central $.5\arcdeg\times.5\arcdeg$, the KS statistic is at the 74\% (40\%) CL
for the distribution of RA (DEC) positions. The test against a gaussian
distribution for the velocities in the same window gives a 24\% CL. (The
skewness (kurtosis) of the velocity distribution, -0.016 (2.8), is at the 47\%
(42\%) CL). As well, the test of Dressler \& Shectman (1988; DS), which
combines the velocity and position information, is also consistent with a
relaxed distribution. Inside the same window we get $\Delta = 58455$ km/s,
where
\begin{equation}
\Delta = \sum_{i=1}^N \delta_i = \sum_{i=1}^N\left((\bar{v_i}-\bar{v})^2
+(\sigma_i-\sigma)^2\right)^{1/2}\ ,
\end{equation}
$\bar{v_i}(\sigma_i)$ is the average velocity (dispersion) among galaxy i and
its ten nearest neighbors, and $\bar{v}(\sigma)$ is the average velocity
(dispersion) among all $N$ galaxies in the window. The significance of this is
obtained by comparing $\Delta$ to the values obtained from 1000 MC reshufflings
of the velocities, which put $\Delta$ at the 19\% CL.

	All these numbers are entirely consistent with a relaxed, isothermal
core in A3266\footnote[1]{Since sampling can introduce noise in the location
of the center of the smoothed map, we have also performed the tests and
calculated CL's by choosing the center at the mean of the positions in a
window. Eq.(1) is replaced by a more complicated expression, but we obtain
similar results.}. These findings are not necessarily in disagreement with
those of Mohr, Fabricant, \& Geller (1993) since the substructure they claim
inside this window in
A3266 (inferred from a systematic shift in X-ray isophote centroids) refers
to the {\it gas} distribution. For example, the gas in a cluster can remain
perturbed much longer than the matter distribution in the case of a merger
(see \cite{RBL93} and the discussion in \cite{PRBHOBH93}).

	Including galaxy positions outside of this core, on the other hand,
we find an increasingly significant deviation from the isothermal distribution,
especially in DEC. The KS test on positions inside a window
$.9\arcdeg\times.9\arcdeg$ yields a CL of 99.96\% (91.1\%) for DEC (RA).
For the entire $1.8\arcdeg\times1.8\arcdeg$ field, we find CL's of 99.9999\%
and 99.8\% for DEC and RA respectively. The DS test gives a $\Delta$ for the
entire field $5.4\sigma$ above the mean of the MC sets, a highly significant
deviation. Thus, the choice of a large field and the measurement of a large
number of velocities is crucial in uncovering this large-scale `anomaly' in
A3266.

	Relaxed dark matter halos are known not to be spherically symmetric,
but are well described by triaxial spheroids (\cite{DC91}, \cite{WQSZ92}).
Therefore, the deviation from our test fit does not immediately imply the
presence of substructure. However, it can readily be seen that the
deviation detected by the RA and DEC KS tests is not due to flattening. In
Fig. 2 we show the cumulative distribution of positions as a function of
position along the RA and DEC axes. It can be seen there that the distribution
of declinations is highly asymmetric, indicating a clear excess of particles
on the north side of A3266. The distribution of RA's is more symmetric,
indicating somewhat of an excess on the east side of A3266. This excess of
galaxies to the N-NE of the cluster can be easily understood in terms of the
`wedge' structure suggested by QRW in a redshift slice through the cluster
(see e.g Fig. 19 in QRW, and Fig. 4(b) below). Indeed, we find that removing
that slice from the data makes the DEC distribution much more symmetric.
However, the possibility that the angular position of galaxies in the slices
in front and behind the cluster (Fig. 15 vs Fig. 16 in QRW) are drawn from
the same distribution is excluded only at the 86\% CL by a KS test.

	By contrast, the DS test suggests a `clumpy' type of substructure.
In Fig. 3(a) we show the distribution of $\delta_i$'s. Given the clear
excess of galaxies with $\delta_i > 600$ km/s, in Fig. 3(b) we show the
spatial distribution of all such galaxies. Several clumps can be seen there,
but no `wedge' of galaxies. Also, despite appearances, most of the $5.4\sigma$
signal is confined to the periphery near and outside the virial radius (marked
by the dotted circle). The CL of $\Delta$ reaches 90\% outside 90\% of the
virial radius, and reaches 96\% at the virial radius. Thus, the signal here
appears entirely consistent with expectations for a relaxed cluster in a
bottom-up hierarchical formation of structure, where subclumps continue to
accrete onto a relaxed core formed early on.

	QRW suggested the `wedge' could be interpreted as a plume of galaxies
resulting from the recent merger of two subunits, with one subunit having
passed through the center from the SW front of the cluster and given rise
to a plume of outflying galaxies. Here we present further evidence in favor
of this hypothesis, and in the next section show that such a feature in the
data (and other characteristics of the data) can indeed be understood as a
result of a recent merger that we simulate by means of a simple N-body
simulation.

	The tidal fields resulting from a merger could be expected to
significantly affect the star formation rate in disk galaxies given
the strong distortions induced in such galaxies when passing through cluster
cores (\cite{D99}).
Therefore emission line galaxies (ELG) can serve as tracers of such an
environment. Many galaxies in the QRW data are ELG, therefore we have
tested if such galaxies indeed trace the plume seen in the redshift slice
of QRW. This would make it plausible that indeed there is a physical
association among those galaxies. In Fig. 4 we plot the position of ELG
(a) below and (b) above the mean velocity. Indeed, there is a striking
difference in their distribution. The distribution seen in Fig. 4(b)
suggests once again that a subclump of galaxies has shot past from the
lower front to the upper back of the cluster. The possibility
that the angular positions of the ELG in front and behind the cluster trace
the same distribution is excluded at the 98.8\% CL by a KS test. We believe
that these results put on much firmer ground the hypothesis that the galaxies
in the plume are physically associated.

%%%%%%%%%%%%%%%%%%%%%%%%%%%%%%%%%%%%%%%%%%%%%%%%%%%%%%%%%%%%%%%%%%%%%%

\section{The Simulations}

	We have performed simple N-body simulations of 1000 particles in order
to investigate the feasibility of the merger scenario proposed by QRW. The
particles start in Hubble expansion within an isolated, uniform density sphere
representing the pre-collapse phase of a cluster. A small amount of angular
momentum is added to represent the angular momentum that would result from
the tidal torquing by nearby condensations. Of course, the collapse and
formation of a cluster is far more complex than what we represent here. Our
aim is simply to explore if a merger giving rise to something like the plume
seen in the data (and with characteristics that would perform similarly under
statistical tests) would happen in this simple, top-hat approximation to the
real collapse of a dark matter halo. We use the N-body code described in
Blumenthal, Faber, Flores, \& Primack (1986), and carry out dissipationless
simulations that represent the evolution of the dark matter.

	In Fig. 5 we show the evolution of a simulation that we analyze in this
section. In this simulation there is a major merger at the center because the
Poisson noise introduced by the discrete realization of the top-hat initial
condition has made the center slightly underdense, therefore the center gets
evacuated (a void forms) and eventually the matter falls in and the merger
occurs. Fig. 5(a) shows (clockwise) the evolution of all the particles, from
the time the
system is near maximum expansion until the time it resembles the situation in
A3266. At this time the velocity distribution of the system closely resembles
that of A3266, and it is the only time at which it does so. In Fig. 5(b) we
show the evolution of the densest groups we can identify prior to the major
meger. The first (top left) panel shows the bottom right panel of Fig. 5(a)
with the group members identified, and follows the evolution (clockwise) of
those groups alone until the time of the bottom left panel in Fig. 5(a). The
groups have mass ratios of 1:2.2:5.5. In the last panel the solid-dot group
is moving away and has developed into a wedge-shape plume that bears a striking
resemblance to the plume uncovered by QRW.

	In the top panels of Fig. 6 we show the distribution of the particles
in velocity slices of the same relative thickness as the 1500 km/s slices in
QRW, both immediately below (left) and above (right) the velocity average. It
can be seen that the plume is still clearly visible in the panel on the right.
The bottom panels show a random sampling of the N-body data, of the size of
the A3266 sample, that is more directly comparable to the A3266 data
(assuming the galaxies closely trace the dark matter distribution). The plume
can still be discerned there, and a KS test on the angular distributions
rejects their compatibility at a very high confidence level.

	The velocity distribution of the particles in the simulation is shown
in Fig. 7 (top panel) together with the same distribution for A3266. They look
remarkably similar, but the difference in peak heights of the dark matter
distribution is not a sampling artifact and suggests that the structure
seen in A3266 might be due to a merger of like-mass subclusters rather than the
heavy-light merger seen in the simulation\footnote[2]{The equal height of the
peaks in A3266 {\it is} a sampling-binning artifact, but even taking this into
account this difference between the simulation and A3266 remains.}. A random
sample from the N-body data, of the same size as
the A3266 sample, is shown in the bottom panel. In this case the KS test on the
N-body distribution excludes the gaussian hypothesis at a higher (99\%) CL than
that for A3266. The skewness (-0.30) excludes the gaussian hypothesis at the
99\% CL (the kurtosis, 2.97, is OK). For most samplings of the N-body
data, however, only the skewness test rejects the gaussian hypothesis with
high significance. Obviously, the skewness test picks up the intrinsic
asymmetry due to the unequal mass nature of the N-body merger. Thus, for an
equal mass merger the results would resemble more closely those for A3266.

	Finally, the DS test on 317-particle samples gives $\Delta$'s
$\sim 5\sigma$ above the mean of the MC sets. As in the case of A3266, the
spatial distribution of the particles with high $\delta_i$'s does not trace
the plume seen in the redshift slice above the N-body cluster's mean velocity.
This is shown in Fig. 8 for two typical samples. Also, the inner core (half
the size the window shown in Fig. 8) is entirely consistent with a relaxed
distribution.

%%%%%%%%%%%%%%%%%%%%%%%%%%%%%%%%%%%%%%%%%%%%%%%%%%%%%%%%%%%%%%%%%%%%%%

\section{Discussion and Conclusions}

	The simulations and the analysis of the data that we have carried out
in the preceeding sections clearly show that the data of QRW on A3266 can be
well explained as the result of a relatively recent, major merger at the core
of A3266. If we fit the x-axis velocity dispersion and the virial mass of the
simulation to the values for A3266, $1161$ km/s and $5\times10^{15}M_{\odot}$
respectively, then we find that the time ellapsed since the cores of the
two massive subclumps roughly coincided is $\sim 2\times10^9$ years
\footnote[3]{We have, self-consistently, fit the dispersion and mass within a
projected window that we then ensure does correspond to
$1.8\arcdeg\times1.8\arcdeg$ at the redshift of A3266.}. We also find that in
this case the true mass has been overestimated by about 70\%. This large a
factor is to be expected in this kind of situation (see \cite{PRBB96}).

	There are many questions that our analysis is not able to address.
Foremost among them is the relation of the merger we claim here, and the
merger claimed by Mohr, Fabricant, \& Geller (1993). Their claim for evidence
of a merger was based on their evidence that the cluster is not relaxed and
evidence by Teague, Carter, \& Gray (1990) of an E-W alignment preference for
galaxies at the extremes of the velocity distribution. However, QRW (as well
others; see QRW) have noticed many discrepancies in velocities deemed
uncertain by Teague, Carter, \& Gray (1990), and it can be seen in Fig. 18 and
20 of QRW that there is no E-W enhancement of such galaxies. Furthermore,
Mohr, Fabricant, \& Geller (1993) point out the agreement in morphology
(elongated) and orientation of the smoothed X-ray intensity and galaxy density
contours, as well as the presence of a secondary peak in the galaxy density
map, as further evidence for substructure in A3266. However, we estimate that
the galaxy map results could be the result of sparse sampling of an otherwise
smooth, isothermal distribution in about 20\% of cases. We show an example in
Fig. 9. Naturally, a hydrodynamical simulation of a major merger like the one
we have advocated here could better test if the X-ray and optical data are all
consistent with such a merger.

	Another issue that we have not been able to address is that of the
dumb-bell system in A3266, most likely itself the result of the merger we
have discussed here. Our simulations do not have nearly the level of resolution
that would be needed to explore if the dumb-bell parameters (separation,
relative velocity, and orientation in the sky) could be explained by the
merger hypothesis. Simulations in the style of those discussed by Dubinski
(1999) would be able to address these question. Here we just note that it is
perhaps significant that the dumb-bell is not centered on the density map,
Fig. 1(a), exactly what would be expected if the system had been formed in a
recent meger. There are many examples of cD galaxies that do not sit at the
bottom of cluster potential wells, precisely in clusters that show evidence of
a recent merger (\cite{B94}).

	There is much observational follow up work that can be suggested to
further test our merger hypothesis. First, the nature of the merger in A3266
that we advocate here, nearly along
the line of sight, makes the core of A3266 an ideal target for a search of weak
gravitational lensing. The redshift of A3266 is low, but there are examples of
weak lensing at this low a redshift (\cite{CKH98}). In this case the cluster is
very massive, and the elongated mass distribution resulting from the merger
will further enhance the weak lensing relative to a spherical cluster with
the same velocity dispersion. We plan to carry out such observations in the
near future. Second, a detailed morphological study of the galaxies in A3266
could help further verify the reality of the merger we have advocated here.
For example, spirals and elliptical might have characteristically different
spatial distributions as a result of type segregation in the pre-merger
subclusters and due to the different effect of tidal fields in the inner and
outer parts of the merging subclusters. We also plan to carry out such a study
in the near future.

	We have carefully analyzed the optical data available on A3266, and
interpreted it with the help of simple N-body simulations to conclude that
there is good evidence in the data that a major merger of comparable-mass
components has occurred relatively recently in this cluster. This analysis
has required a wide-area coverage in the cluster, as well as a large number
of galaxy spectra in order to uncover the large-scale plume of galaxies that
we have advocated here to be a telltale sign of a recent, major merger.
This opens the prospect that under similar scrutiny other, perhaps many? (of
the 20\% we have mentioned),
apparently relaxed clusters might be discovered to actually be undergoing a
major merger. The frequency with which such process is seen to occur in
nearby clusters might then tell us about the underlying cosmogony generating
them.

%%%%%%%%%%%%%%%%%%%%%%%%%%%%%%%%%%%%%%%%%%%%%%%%%%%%%%%%%%%%%%%%%%%%%%
\acknowledgements 
This work has been supported by an NSF grant and by a Research
Board award at UM--St.Louis, and by FONDECYT grants 8970009 and 7960004 at
PUC. RF would like
to acknowledge the hospitality of the Physics Department at UC--Santa Cruz
where some of this work was carried out. HQ was partially supported
by the award of a Presidential Chair in Science. This research has made use
of NASA's Extragalactic Database, ADS Abstract Service, and the Digitized
Sky Survey at STScI.

%%%%%%%%%%%%%%%%%%%%%%%%%%%%%%%%%%%%%%%%%%%%%%%%%%%%%%%%%%%%%%%%%%%%%%

%%%%%%%%%%%%%%%%%%%%%%%%%%%%%%%%%%%%%%%%%%%%%%%%%%%%%%%%%%%%%%%%%

\begin{figure}

\caption{Smoothed galaxy density maps. The data for A3266 are shown in
(a) smoothed with a gaussian window of $\sigma = 1.8\arcmin$, corresponding
to $\sim 35h^{-1}$ kpc at the redshift of A3266. This is of the order of the
soft core radius seen in weak lensing studies of cluster mass distributions
(see \cite{TKD99}). The filled dots mark the location of the dumb-bell
components. The center of the plot is at RA = 04:30:30.7,
DEC = $-$61:33:25 [1950]. A MC realization of the data is shown in (b),
smoothed on the same scale. All contours shown are spaced at 10\% intervals,
starting at 15\%, of the maximum density.}

\caption{Cumulative distribution of DEC (solid) and RA (dotted) in the
entire $1.8\arcdeg\times1.8\arcdeg$ field covering A3266. The expectation
for an isothermal sphere is shown by the dashed line.}

\caption{(a) Distribution of deviations $\delta_i$ (eq. 2) for all the
galaxies in A3266 (solid line), and the average distribution for the 1000 MC
reshufflings of their velocities (dashed line). (b) Position in the sky of
all the galaxies in A3266 with $\delta_i > 600$ km/s. The virial radius of
the cluster is indicated by the dotted circle.}

\caption{Position in the sky of the galaxies in A3266 with velocities
(a) below and (b) above the mean for the cluster. ELG are shown by
solid squares, and others by crosses.}

\caption{Major merger in N-body simulation. In (a) we show the time evolution
(clockwise from top left) of the system as a whole. The panels have the same
size and show the projected distribution of particles in the simulation onto
the plane perpendicular to the system's angular momentum. The system is near
maximum expansion in the first frame, and the last frame shown is the time at
which the particle velocity distribution is similar to that of A3266. The
softening parameter is 1/160 of the size of the frames. In (b) we show the
time evolution of the three densest groups, identified at the time shown in
the bottom right panel in (a). The panels have half the size shown in (a).
The figures show the projected distribution of particles in the groups onto
the plane perpendicular to the x-axis of (a), which points out of the figure.
Only the first panel shows all the particles, with those not in the groups
shown by small symbols. The time is that of the bottom right panel in (a).
Subsequent panels show the evolution (clockwise) of the groups until the time
of the bottom left panel in (a).}

\end{figure}

\clearpage

\begin{figure}

\caption{Velocity slices through the dark matter distribution, perpendicular
to the x-axis of Fig. 5, after the major merger has taken place. The time is
that of the bottom left panel in Fig. 5(a).
All panels show velocity slices of thickness (relative to dispersion)
equivalent to 1500 km/s slices for A3266 (see Fig. 17-20 in QRW). The panels
on the left are slices immediately below the mean, whereas those on the right
are above the mean. The top panels show all the particles, whereas those at
the bottom show those of a random sample of size as the A3266 data set. The
dotted lines mark the angular sector where the distribution of position
angles of the particles in the slice is markedly above that of the slice on
the left. See the discussion in the text for further explanations.}

\caption{Velocity distributions. The top panel shows the distribution of galaxy
velocities in A3266 (solid line) and particle velocities in the simulation
discussed in the text (dotted line). The bins are 200 km/s wide for A3266 and
proportionately (relative to dispersion) thick for the simulation data. The
height of the middle peak has been set equal for comparison. The bottom panel
shows the A3266 data compared to that for particles in a random sample of the
size of the A3266 set. See text for further discussion.}

\caption{The top panels show the position of particles in the velocity slice
above the mean (see Fig. 6) for two random samplings of the size of the A3266
data set. The bottom panels show the positions of the particles
with the highest $\delta_i$'s in the random sample above. The threshold in
$\delta_i$ was chosen in the same manner as the A3266 case (see Fig. 3). The
dotted lines marking the angular sector shown in Fig. 6 have been kept for the
top panels.}

\caption{Smoothed `galaxy' density map of a MC realization of an isothermal
sphere, using eq.(1), with sample size as that of A3266. Contours shown are
spaced at 10\% intervals, starting at 15\%, of the maximum density. See text
for futher explanation.}

\end{figure}

%%%%%%%%%%%%%%%%%%%%%%%%%%%%%%%%%%%%%%%%%%%%%%%%%%%%%%%%%%%%%%%%%%%%%%

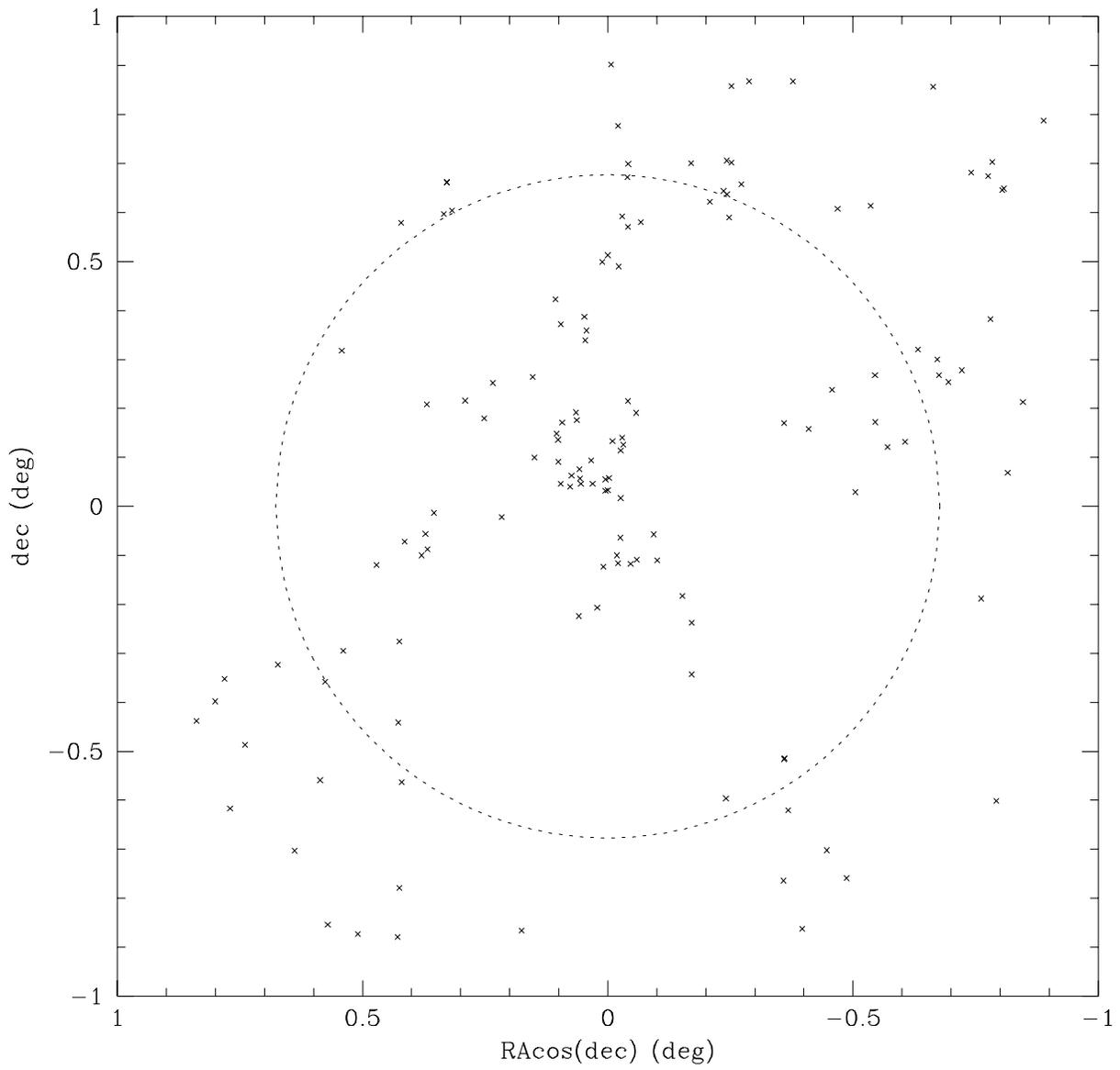
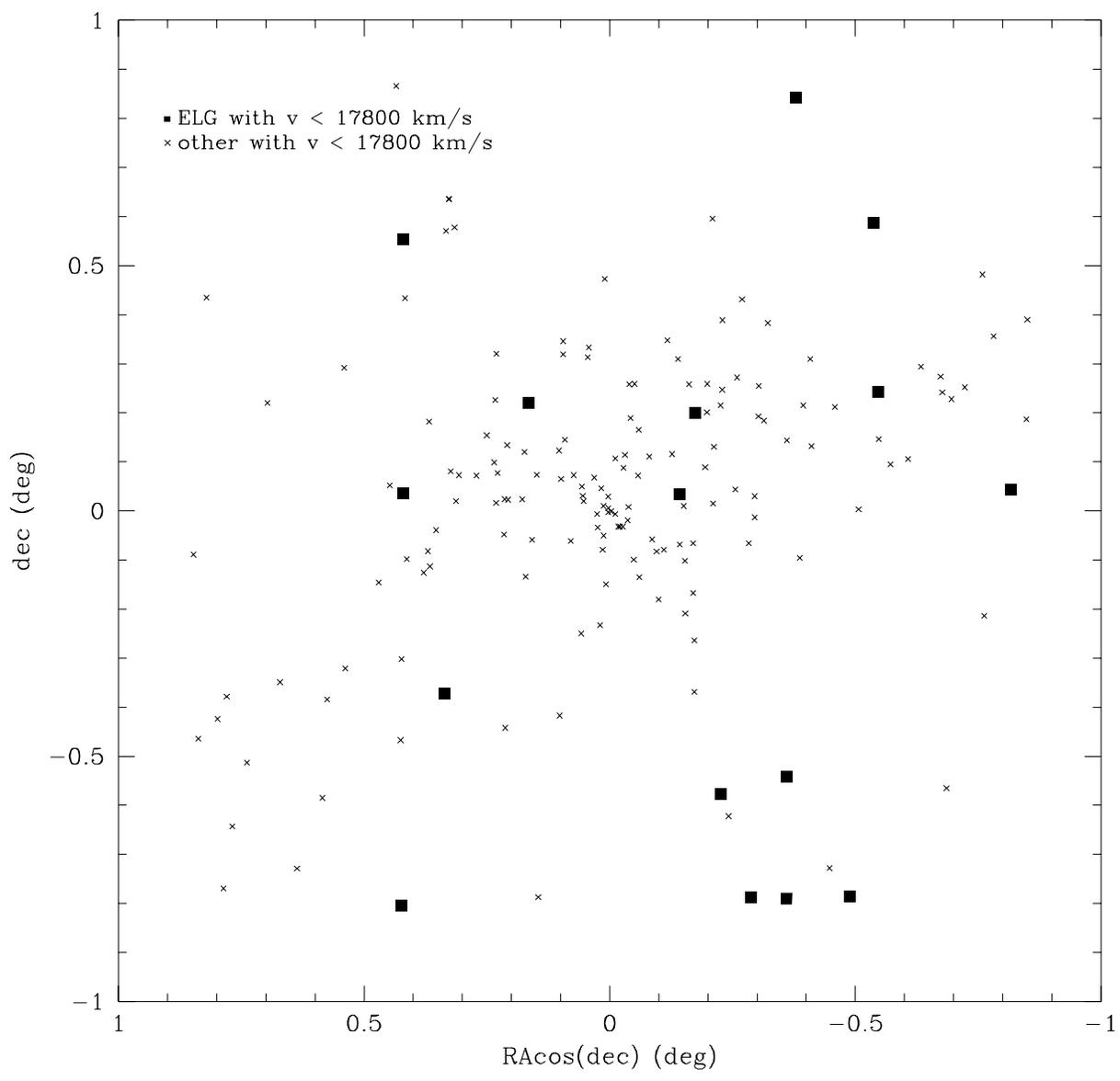
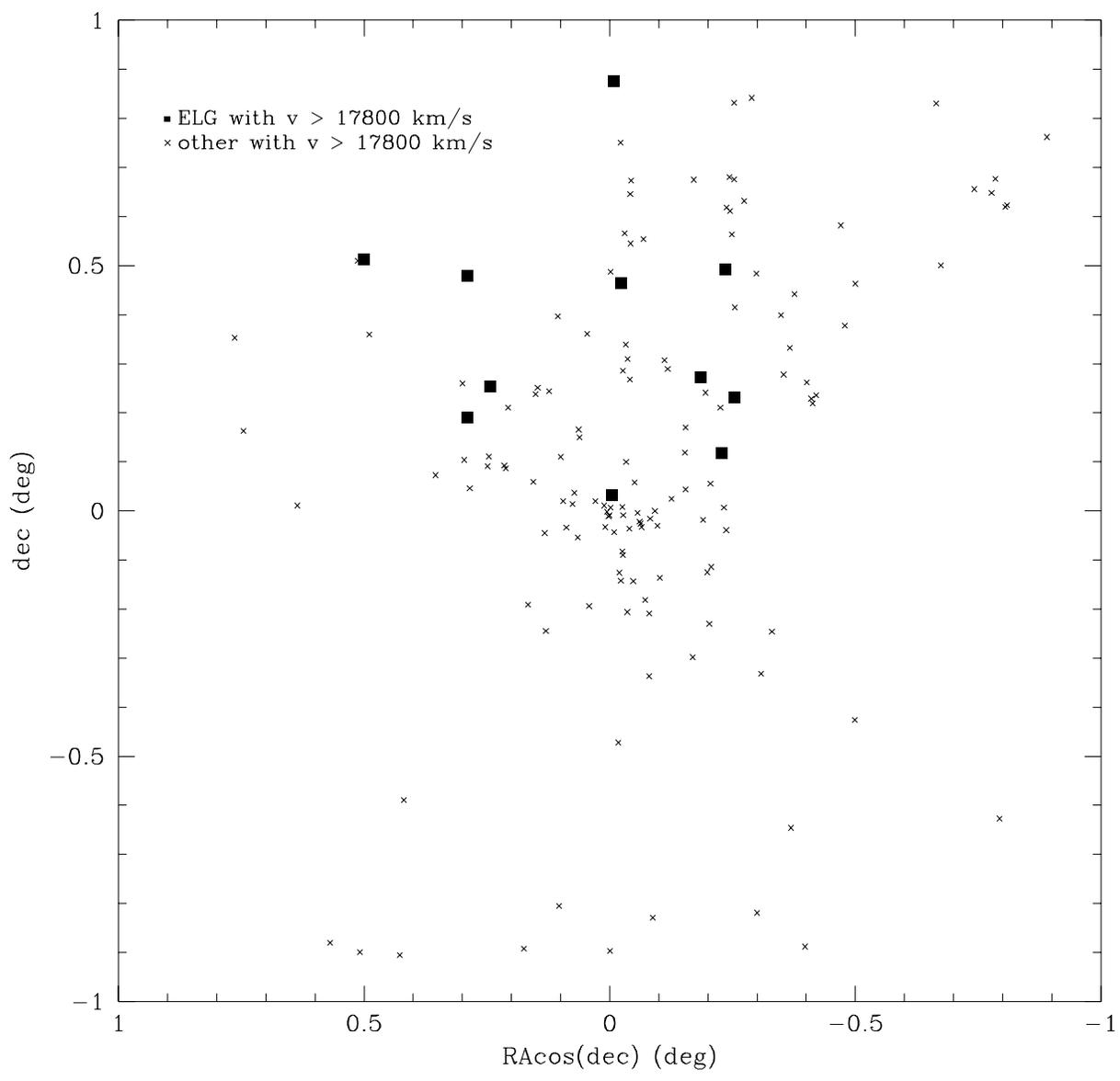
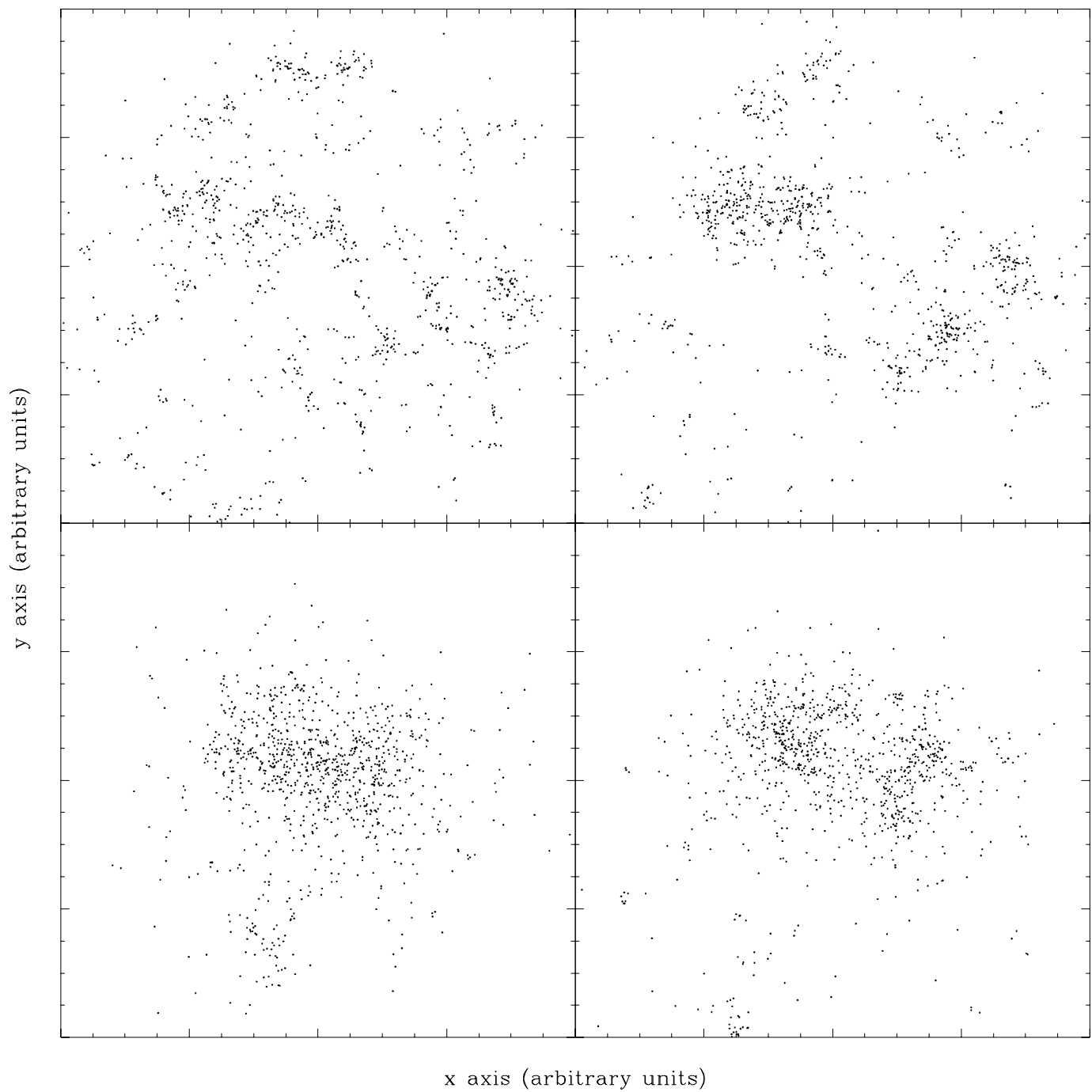
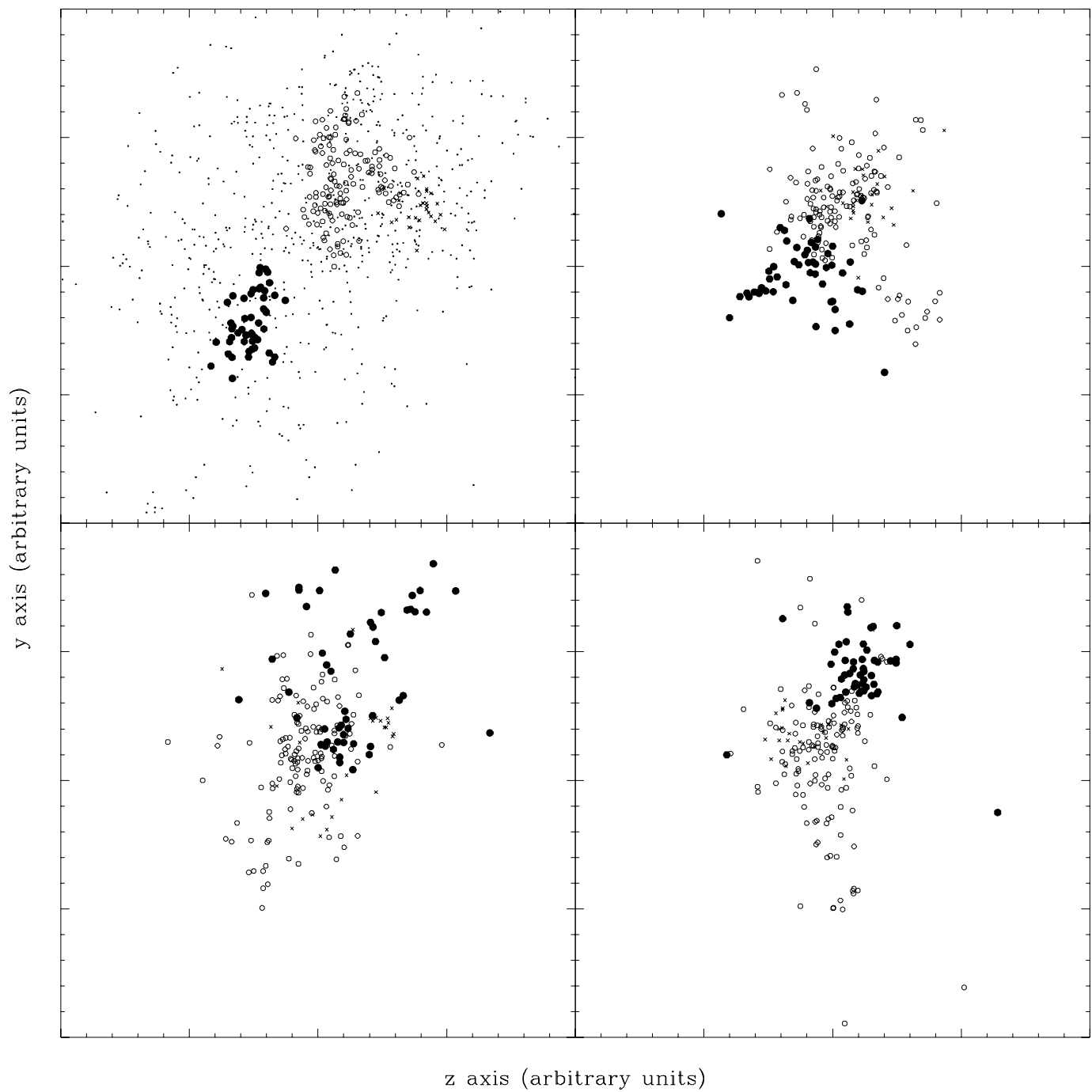
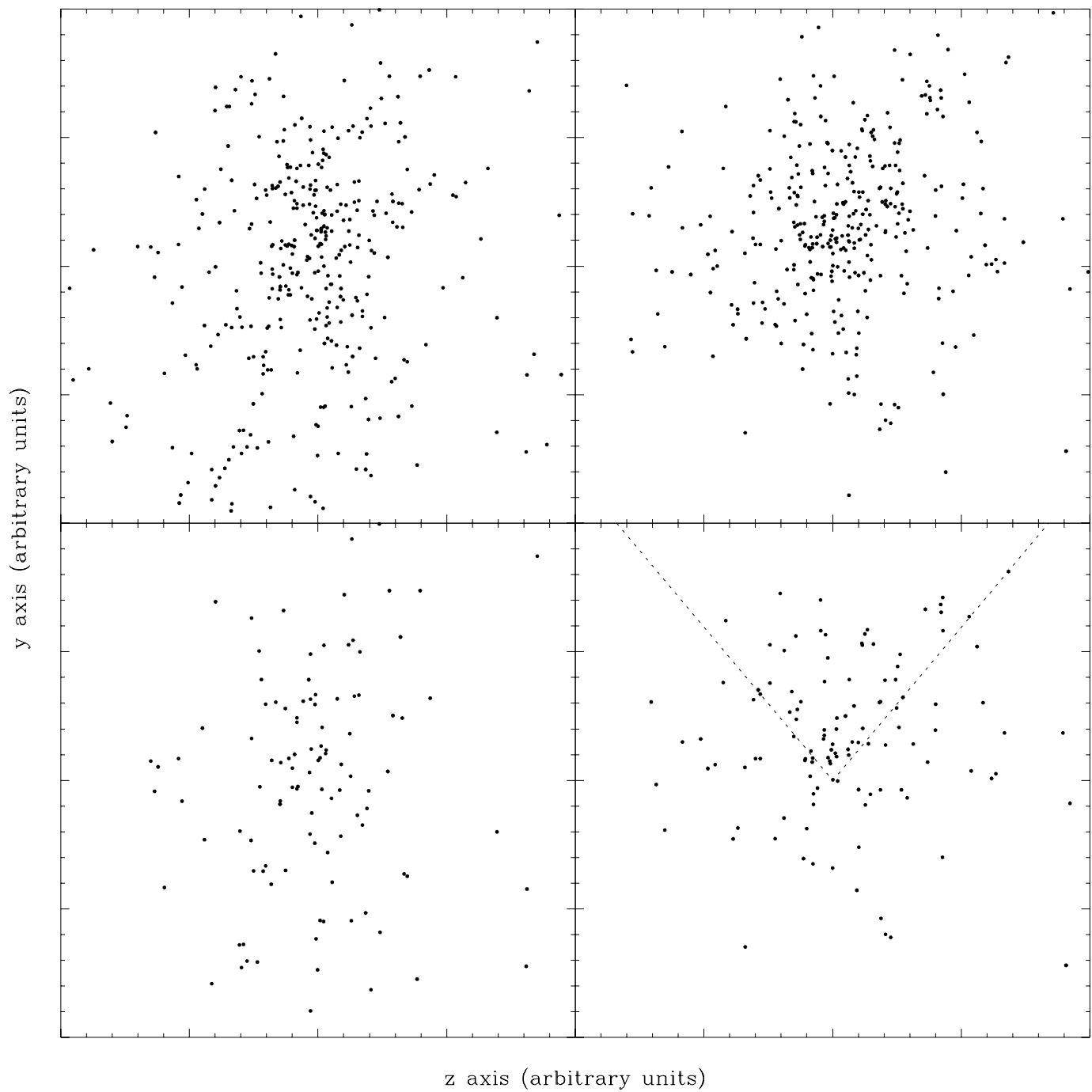
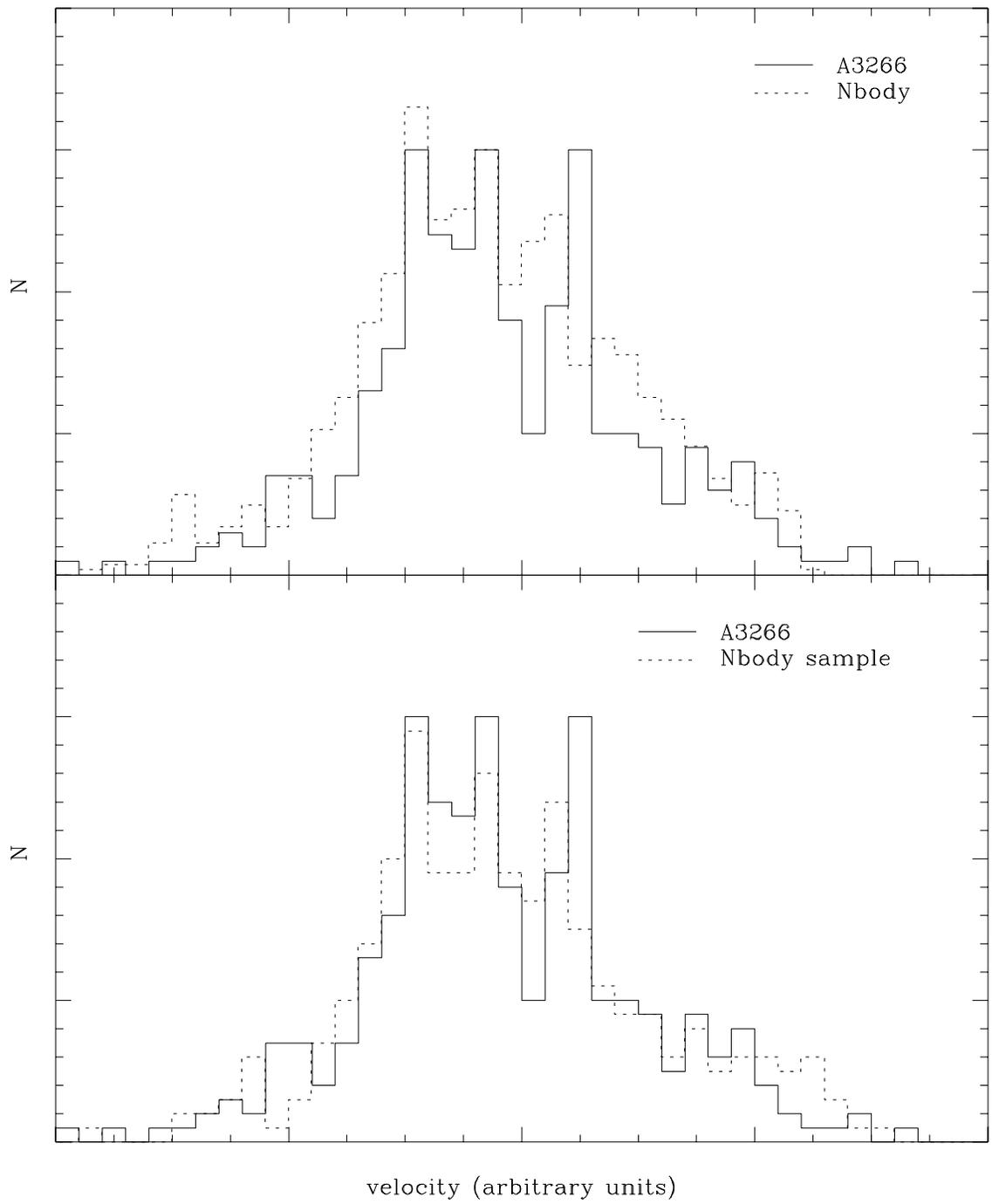
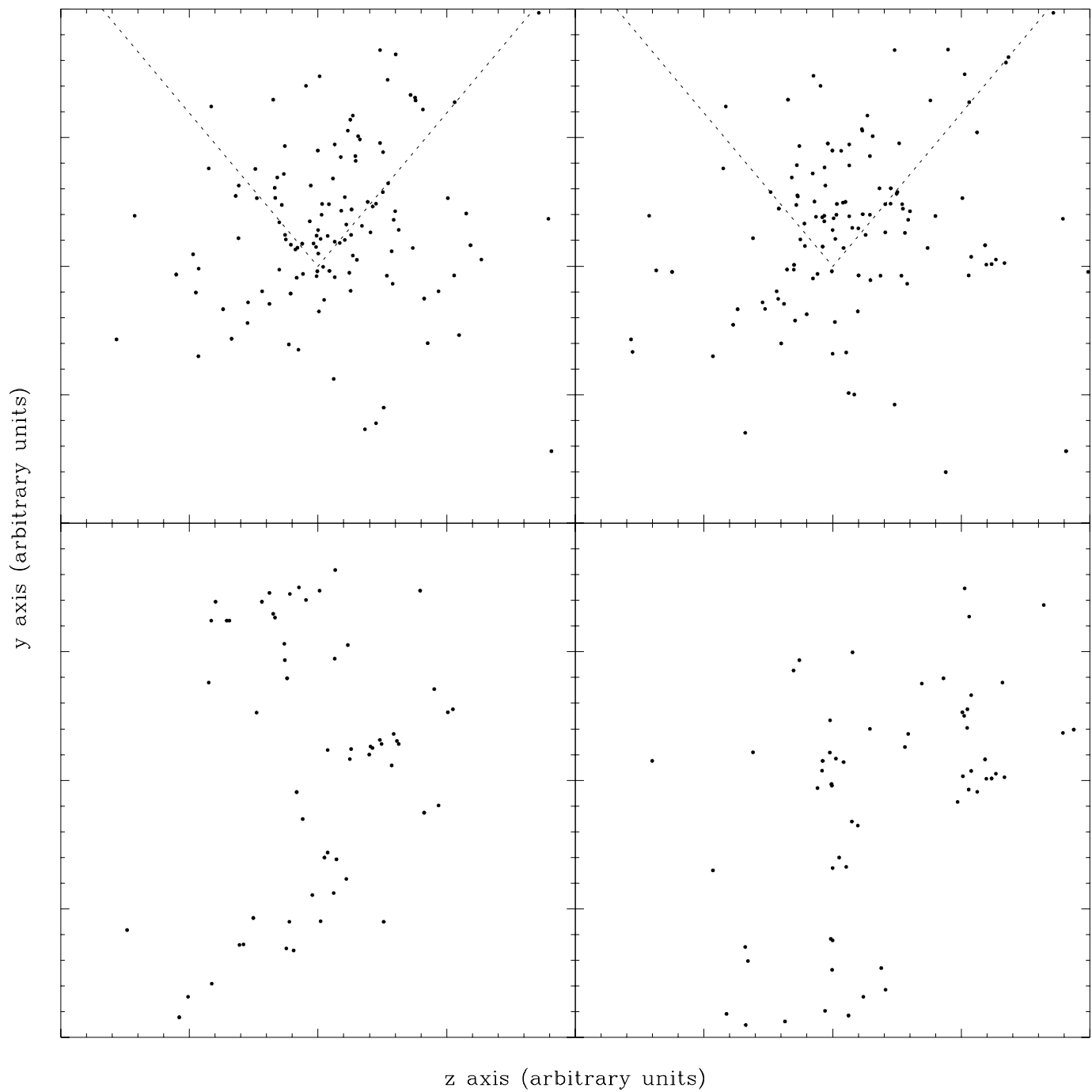
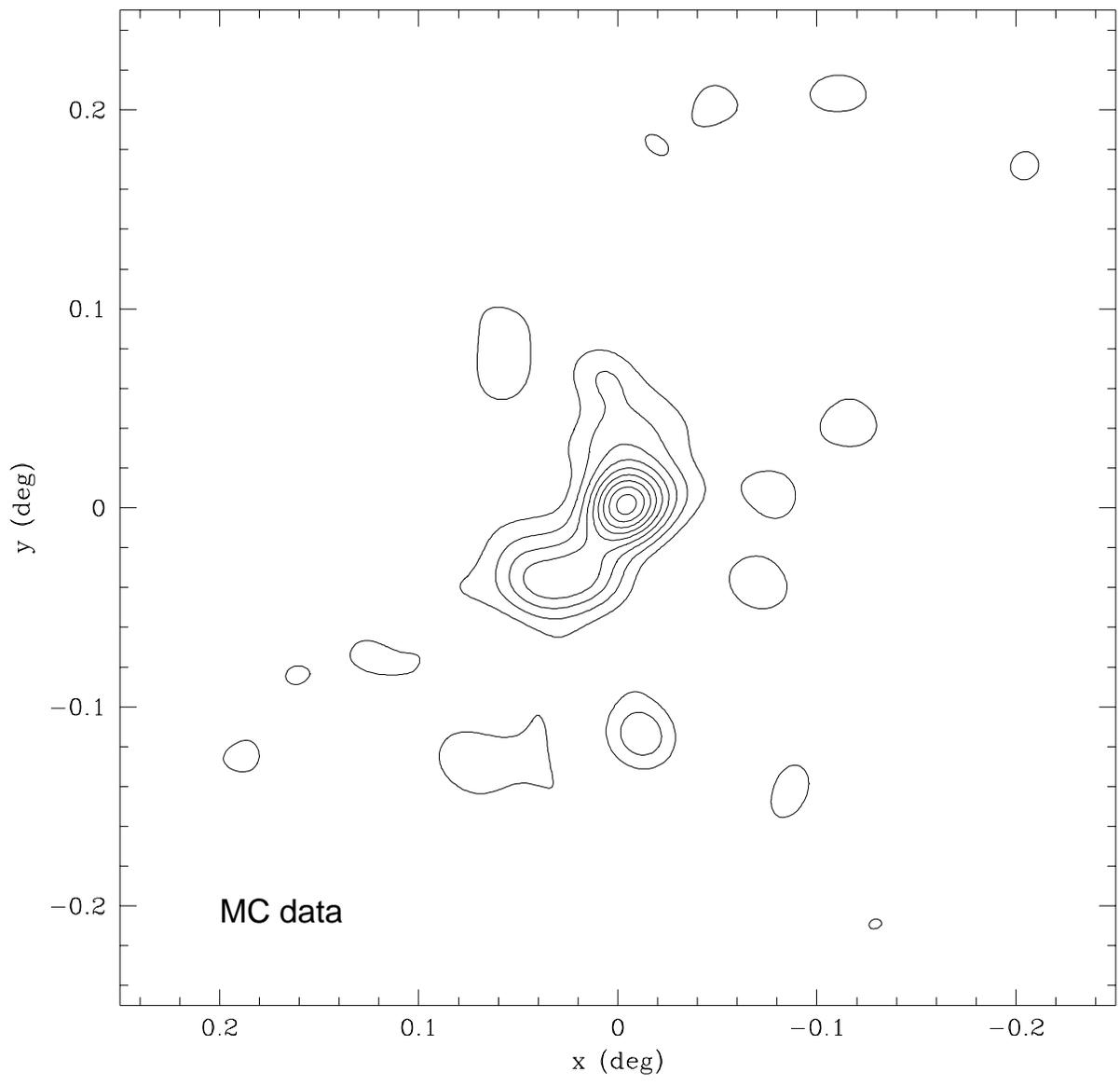
\end{document}